%% file: rmag-new-grb.tex
\documentstyle{article}          
\catcode`@=11

\@addtoreset{equation}{section}
\catcode`@=12

\title{\centerline{%
}\bigskip
\bf Neutrino Absorption: In The Magnetic Field Of GRB In The Fireball Model.}

\author{\bf A.  K. Ganguly$^a$\thanks{aviavi85@rediffmail.com}
\normalsize 
\\
\normalsize
Wintel Communications, Kolkata 700030\\ 
India.\\ 
\normalsize 
}

\date{November 2004}

\begin{document}

\maketitle

\begin{abstract} \noindent\small 
 
TeV or Super TeV neutrinos are expected to originate at the Gamma ray burster
( GRB ) events in the universe. These neutrinos are expected to be produced 
from the photo-meson interaction of the protons  in the GRB environment.
In the usual picture, the protons in the GRB fireball undergo Fermi 
acceleration in the ambient magnetic field of the GRB to very high energy.
These protons then interact with the MeV, KeV or meV photons to produce
Delta particles those subsequently decay to produce very high energy neutrinos.
In this note we focus on the 100 TeV neutrinos produced in a GRB in the 
fireball model. We try to estimate the production point of these multi
 TeV neutrinos from the center. The strength of the Magnetic field there. 
Lastly but importantly modification  in the spectrum because of neutrino 
absorption in the ambient magnetic field. The strength of the ambient 
magnetic field may change the neutrino spectrum. 
\end{abstract}

\section{Introduction}\label{gp}

It is  established now that GRB's are among the most violent energy 
releasing events of the universe. Total energy release in these events
could be equivalent to few solar mass and are seen to be 
distributed isotropically all around the universe.
They are very high redshift objects occurring in nature -- implying
their cosmological origin. Since their discovery, the physics of the 
process that powers these  events is yet to be understood. For some time 
now it has been conjectured that GRBs are source of very high energy 
cosmic rays. Moreover they have also been tipped off as one of the 
source of very high energy neutrino. The energy reachable in the 
GRB mediated events are several orders of magnitude higher than
those available in the laboratory, that is one of the motivations
why there is some interest in the community to study the physics of GRB.
\\ 

\noindent
The origin of high energy particles in a GRB is ascribed to Fermi  mechanism
in the ambient magnetic field in the GRB enviornment. As we discuss later in 
this paper the same magnetic field might be responsible for modifying the 
$\nu$ spectra. This requires good understanding of GRB physics as well as,
 careful evaluation of GRB parameters like temperature, Magnetic field, jet
 Lorentz Boost factor, matter density in the rest frame of the jet etc. We 
discuss them below one after the other.\\

\noindent
As already mentioned, GRB's are very high redshift objects, implying,
they must have happened during the  time of formation of the first generation  
stars of our universe.
They have  certain distinct features, such as, the rapid variability 
time ($10^{-2}$ Sec), the non-thermal nature of the spectra, the hardness 
of the spectra ($\sim$ 100 MeV ). The  existence of large energy in 
the power law tails of the GRB spectra--- above threshold gives rise to 
whats usually referred in the literature as {\it{ compactness problem}}
\cite{compact}. The naive estimate of the source size from the variability
of the spectra runs in to problem. This is principally because of the fact, 
that a stationary compact object with an energy release, of the order of 
$10^{52}$ erg/sec would produce an optically thick environment; however
the fact that one observes non-thermal spectra, gave rise to what is called 
{\it{Compactness Problem Of GRB}}.    
The time variability and hardness of the spectra together suggest
the source to be extremely compact in size  ( $r_o$ $\sim 10^7$ cm ), 
from which  a highly energetic jet moves outward  around the rotational 
axis of the object.Actually the unusually hard and 
 non-thermal nature of the photon spectra  suggest that the jet must be moving 
towards the observer with high Lorentz boost $\Gamma$, whose magnitude can 
vary between  several hundreds to few tens of thousands. \\

\noindent
Inside the moving jet, the medium has appropriate optical thickness 
to last the radiation in the jet up to the variability time ( of the jet).
That could vary between  $1 \mbox{~to~} 10^{-4}$ sec. To have the jet 
optically thick to ( One to two MeV photons ), it has to have some baryonic 
contamination in -- composed of p and n, \cite{mes}.
They are accelerated to very high energy in the shock and ambient magnetic 
field of the GRB through Fermi mechanism and subsequently  would produce 
pions through photo-meson or internal collision processes. 
These  pions finally decay to high energy neutrinos, electrons  
and positrons. In this note we would be focusing mainly on neutrinos.\\

\noindent

The neutrinos produced in GRB, are emitted at various stages of the jet 
evolution with varying energy. Starting from low energy neutrinos at the 
beginning to extremely high energy neutrino energy at the end. Typically 
GeV neutrinos
are expected to be produced at the initial stage, followed by 5 TeV or more
and then $\ge$ 100 TeV ones.\\

\noindent
Neutrino properties are known to be sensitive to ambient Electro-magnetic 
field  and matter field \cite{ganguly}. In particular there are several
 decay channels, forbidden in vacuum but are allowed in a background magnetic 
field. Actually these are the motivating reasons to look into aspects 
of the magnetic field structure, medium properties; at  locations
where neutrinos of specific energy would originate. 
With this purpose in mind, in this note we  would estimate the radial distance 
form the center of the star ($r_{\pi}$) where the energetic pions form to 
decay into 100 TeV neutrinos. We then go ahead to estimate the magnetic field 
there, followed by estimates of magnetic field induced neutrino decay.\\

\subsection{Neutrino Production In GRB.}

\noindent 
As has been mentioned in the beginning, in this note we investigate ultra high energy neutrino production
and their interactions in the ambient medium and the possibility
of change in the spectrum because of that. 
The reaction dynamics is as follows, pions first decay to muon and muon 
type neutrino, i.e., 
\begin{eqnarray}
\pi^{+} \to 
\mu^{+} + \nu_{\mu},
\label{react1}
\end{eqnarray}
followed by the muons  decaying to, 
\begin{eqnarray}
\mu^{+} 
\to e^{+}+\nu_{e}+\bar{\nu}_{\mu}. 
\label{react2}
\end{eqnarray} 
From  Eq.(\ref{react1}) and (\ref{react2}) its clear that,  $\nu_{\mu} \& 
{\bar\nu}_{\mu}$
emission takes place from two different reactions.
The first emission would be taking place within a radius ( $r_{\pi}$, the 
pionosphere radius)
within which the pions produced from photo meson interaction would decay.
The other region being the location where the energetic muons decay (usually 
referred in literature as $r_{d}$).
For a given $r_d$ one can derive the energy of the muon anti-neutrinos observed on earth 
( $\epsilon^{ob}_{\bar{\nu}\mu}$ ) or vice versa. The muon anti-neutrino in 
Eq. (\ref{react2}) is an end product of a three body decay, therefore the
energy carried by the $\bar{\nu}_{\mu}$, in the rest frame of the muon--
 would be $\frac{1}{3}m_{\mu}$ (roughly!). Where $m_{\mu}$ is the rest mass 
of the muon. Hence the anti-neutrino energy as observed on earth would be 
given by,
%
\begin{eqnarray}
\epsilon^{ob}_{\bar{\nu}\mu}=\frac{1}{3}m_{\mu} \Gamma_{\mu}\Gamma_{h}, 
\mbox{~ In units of 
\rlap/h=c=1.~}
\label{1/3}
\end{eqnarray}
%
Where $\Gamma_{h}$ is the Lorentz boost of the wind with respect to an 
observer on earth. On  the other  hand, if the muons with half life 
$t_{\mu}$ originate in the jet wind of the GRB ( extending up to distance 
$r_d$), then it follows from there that,
%
\begin{eqnarray}
r_d=\Gamma_{\mu} \Gamma_{h}t_{\mu}.
\label{rd}
\end{eqnarray}
%
From relations Eq.(\ref{1/3}) and (\ref{rd}) one can relate the observed
anti-neutrino energy with that of the wind dynamical radius, i.e.,
%
\begin{eqnarray}
\epsilon^{ob}_{\bar{\nu}\mu} = \frac{m_{\mu} r_d}{3 t_{\mu}}.
\label{evr}
\end{eqnarray}\\
%
\noindent
For the muon type neutrinos, produced in a frame where the pion is at rest,
 the available kinetic energy is roughly close to 
$(m_{\pi} - m_{\mu})=$ 34 MeV and the resulting muon is non-relativistic. 
So even if most of the energy is carried by the $\nu_{\mu}$ it would not
be very different from the $\bar{\nu_{\mu}}$ or the $\nu_{e}$, i.e 
$ \sim \frac{1}{3} m_{\mu} $. Although their point of origin in the GRB
environment would vary ( because of difference in lifetime of pion or muon),
hence change in the Lorentz Boost, however given the detector
efficiency, that probably would not be detectable. Therefore we would consider
the point of their origin to be same.\\   

\noindent
We would like to note, in absence of magnetic field the ratio of number 
density of neutrinos ($n_{\nu_{e}} n_{\nu{\mu}}$) would be 1:1, at the production
point. However in a magnetic background new decay channels, e.g., $\nu \to
\nu l^{+} l^{-}$ opens up with nonzero probability $\alpha$. In general
$\alpha$ depends of lepton mass $m_l$ their energy $E_{\nu}$, strength 
of Magnetic field B etc. Since these many of these quantities are flavor
dependent, implying different absorption rate for different neutrino flavors.
And to find out the absorption rates we need to estimate the parameters
like Strength of Magnetic Field at a particular location. The energy of the
neutrinos being produced there etc. In view of this this has been estimated
in this paper, following the fireball model of \cite{mes}.

\indent
To estimate various physical parameters of a GRB, we would stick to
the fireball model of GRB ( successful ) and try to refrain from making 
any explicit assumption about the progenitor. 
The  postulates and assumptions we make, would be  based  mostly on 
\cite{mes}, \cite{wax-mes}, \cite{bah-mes} and \cite{wax-bah}.\\

\noindent
The organization of this document is as follows, in section II, following
 fireball model  we discuss the physics of GRB. Section III is devoted to 
estimating various parameters, and lastly section IV would deal with
details of absorption rate and the asymmetry in  number density of
$\nu_e$ and $\nu_{\mu}$ for some specific energy $\ge 100$ TeV. we conclude
by discussion and some numerical estimates of asymmetry. 
\section{Physics Of GRB}
\noindent
Numerical simulations based studies reveal that, the  jet produced 
at the central engine of the GRB moves along the rotation axis with a Lorentz 
factor (Lf) $\Gamma$.
After launch.jet consists of shocked jet material and shocked stellar 
materials. Initially, the jet heats up the region 
where it interacts with the stellar material, till the pressure in this zone
 becomes comparable to the same in the jet. Following notations of 
\cite{bet} region ahead of  shock frame is called up stream, the one behind 
the shock called down stream. As the pressure in the up stream termination
shock grows material flows into the jet creating a reverse shock ( Lorentz
factor $\Gamma_R$). Subsequently material in the shocked region down stream
 (jet head) move out ward with Lorentz factor $\Gamma_h$. The synchrotron
and inverse Compton is believed to diffuse in, thermals and power the 
movement of the jet head. 
The isotropic luminosity of the GRB usually is related with the 
temperature of the medium in the jet head.In the rest frame of the 
jet ( of Lf, $\Gamma_h$ ) the retarded jet moves with a 
Lorentz factor $\frac{\Gamma_j}{2\Gamma_h}$ \cite{wax-mes}.

After injection, during its passage through C/O/He core jet head moves with 
subreltivistic velocity. However beyond He core there is a significant
drop in the density of the stellar material (envelope)i.e., $\rho \le 10^{-7}$
gm/cc. and the jet can accelerate. As will be discussed below, during this 
phase $\Gamma_h \propto \frac{r}{r_o}$, where $r$ and $r_o$ are the radial
distance at the point of concern and the injection point. The jet accelerates 
till it's L.f  reaches a saturation value. Further out it follows a self-similar motion and decelerates while interacting with the (ISM);we note here that
baryonic components (mainly protons) of the jet are accelerated to high 
energy in the reverse shock and these high energy protons interacting
with the thermal photons produce neutrinos.\\

\noindent
Near the outer envelope, jet is believed to be dissipation-less hence 
the jet fluid in GRB can be collision-less. It behaves like an ideal 
Magneto Hydro Dynamic (MHD) fluid. 
As a result it follows that the entropy 
and the energy of the fluid component can be held constant.\@ Conservation 
of entropy and  energy implies \cite{waxmanl},
\begin{eqnarray}
r^2\Gamma(r)r_o T^{3}_{\gamma}(r) = \mbox{constant}.
\label{cons}
\end{eqnarray}
and 
\begin{eqnarray}
r^2\Gamma(r) r_o \Gamma(r) T^{4}_{\gamma}(r) = \mbox{constant}.
\label{cone}
\end{eqnarray}
In the equations above, $r_o$ stands for radius at the injection point 
and its usually taken as $r_o=10^{7}cm$.
We would like to point out that, $\Gamma(r)$ the position dependent
Lorentz factor, from now on, would be replaced by $\Gamma_h$ unless 
explicitly mentioned.\@ From the estimates of Luminosity in terms of 
temperature one obtains, the estimate of the temperature at the 
injection point turns out to be,
\begin{eqnarray}
T_{o}=1.2 L^{1/4}_{52}r^{-1/2}_{o7} \mbox{Mev}.  
\end{eqnarray}
With $L_{52}=10^{-52}L_{iso}\mbox{\@ erg/sec}$ and $r_o=10^{-7}r_{07} \mbox{\@ cm}$. 
Assuming, $\Gamma(r)|_{r=r_o}=O(1)$, it is possible to estimate the rhs
of eqns. (\ref{cons}) and (\ref{cone}). They turn out to be,
\begin{eqnarray}
r^2\Gamma(r)r_o T^{3}_{\gamma}(r) ~~\sim ~~2.1 \times 10^{53} L^{3/4}_{52}.
\label{cons-1}
\end{eqnarray}
and 
\begin{eqnarray}
r^2\Gamma(r) r_o \Gamma(r) T^{4}_{\gamma}(r)~~\sim~~ 2.5 \times 10^{53}
L_{52} \mbox{MeV}.
\label{cone-1}
\end{eqnarray}
As before unless explicitly mentioned we would suppress the explicit position
dependence of temperature $T(r)$. Relations (\ref{cons-1}), 
(\ref{cone-1}) imply,
\begin{eqnarray}
\Gamma_h \propto r \mbox{\@.} 
\label{gamma-pro-r}
\end{eqnarray}

%
%
\noindent 
Assuming the presence of only magnetic field,  ($\left( eB \right )$, in 
the rest frame of the plasma.), using dissipation-less fluid equations, it 
is possible to derive an exact relation,
\begin{equation}    
\left(\frac{eB}{2}\right)^2=  n_j m_p.
\label{equipart}
\end{equation}
Where, we assumed the medium to be composed of heaviest 
charged particle available in the plasma, namely protons . Their number  
per unit volume in jet rest frame is defined as  $n_j$. This is justified 
because the contribution from electrons would be far too less.
Further more if we assume them to carry the isotropic energy ( luminosity ),
$\mbox{L ergs/sec}$, then the number density of protons, at some distance 
$\rm{r}$ ( to an observer on earth ) from the center of the GRB ( moving with 
Lf, $\Gamma_j$ ) towards earth would be \cite{wax-bah},
\begin{eqnarray}
n_{j}= \frac{\rm{L}}{4 \pi r^2 \Gamma^2_j m_p}.
\label{pd2}
\end{eqnarray}
For a locally charge neutral plasma, Eq.(\ref{pd2}) would also provide the 
number density of electrons. 
\\

\noindent
The physics of the shock front is described by Hugoniot equations.They 
 balance various thermodynamic quantities like, energy, pressure,
etc., across the discontinuity of the shock front. For instance
equating the pressure between the
forward and backward shock, (with density of material $\rho$ in units of
gm cm$^{-3}$) it has been shown \cite{wax-mes} that ,
%
\begin{eqnarray}
\Gamma^{4}_{h} &=& \frac{\Gamma^{2}_{j}}{4 \rho} n_{j} m_{p} 
\nonumber \\
&=& \frac{L}{16\pi r^2 \rho}.
\label{gammah}
\end{eqnarray}

Following \cite{mackee} near the outer envelope one can parametrize 
$\rho= \rho_{*}\left(\frac{R}{r} -1 \right)^3$ with $\rho_{*} \le 10^-7$gm/cc.
Using this parametrization, it turns out that,
\begin{eqnarray}
\Gamma_h = \frac{Lr}{16\pi \rho_{*} \left(R - r \right)^3}
\end{eqnarray}
which matches with the suggested form $\Gamma_h \propto r$, if one expands
the denominator in powers or $\frac{r}{R}$, for $r < R$ and retains the 
leading order term. 
%
%
\section{ Parameters }\label{para}
\subsection{Temperature Of the plasma}
%
\noindent
In this section we would try to estimate the temperature, that the shocked
plasma jet moves with.\@ Since photons are believed \cite{mes} to be radiated 
from the shock heated  plasma jet, moving with Lf $\Gamma_h$ then all the 
radiated photons from the GRB would account for its total luminosity 
$\rm{L_{iso}}$. Using the emissivity-temperature
relationship for pure photon gas it is possible to establish the 
temperature of the emitted radiation. The same under this assumption
 turns out to be,
\begin{eqnarray}
T_r \sim 10\left( \frac{\rm{L_{iso}}}{r^{2} \Gamma^2_h} \right)^{1/4},
\label{temp}
\end{eqnarray}

However this relation is bound to get multiplied by correction factors
($<1$) if the photons constitute a part of the total luminosity L. However 
to keep the physics simple we assume the photons to carry the total 
energy of the GRB. 
We conclude this section by noting that, the number 
density of the photons at this temperature
would be given by.
\begin{eqnarray} 
n_{\gamma}= \rm{T}^{3}_{r}.
\label{phnd}
\end{eqnarray}
which essentially follows the laws of pure photon gas. One can use eqn. 
(\ref{temp}) to express photon number density in terms of total luminosity
L and radial depth from the center. 
\subsection{$r_{\pi}$: Pion Formation Radius }\label{pir}

\noindent
In previous sections we have estimated the various parameters related to the 
production of high energy protons as well as the approximate temperature 
profile of the shocked plasma head. In view of these estimates
it is  appropriate to estimate the the distance from the origin where
these protons ( of energy $\epsilon_p$ ) would scatter off the thermal 
photons to undergo $\Delta$ formation.  
 In the jet head, the protons at the resonant energy would scatter off 
thermal photons to produce $\Delta$ as long as,at the rest frame of the jet,
 the following energy condition, in obvious notation, is given by, 
\begin{eqnarray}
\epsilon_{\gamma} \epsilon_{p} = 0.2 \rm{GeV}^2.
\label{ph-pr-condn}
\end{eqnarray}
Before proceeding further we note that, the reaction rate 
$ R$ for $p + \gamma \to \Delta$ would be given by
\begin{eqnarray} 
R = n_{\gamma} n_{j} \sigma_{p,\gamma \to \Delta} 
\left[1- cos \left(\theta\right)\right]
 \sim  n_{\gamma} n_{j} \sigma_{p,\gamma \to \Delta}, 
\end{eqnarray}
Here we have assumed $\theta = 90^{o}$. Its appropriate to note that, in 
order to have resonant photon proton  scattering to $\Delta$, inside a 
GRB shock cavity, number of protons ($n_j$) in unit volume should 
be equal  to  $R \times \delta t$; with $\delta t$ being variability time 
for the GRB, this leads to the condition,
\begin{equation}
n_{\gamma}\sigma_{p\gamma \to \Delta} \delta t = 1.
\label{rescond}
\end{equation}
In eqn.(\ref{rescond} ) above,
 $\sigma_{p \gamma \to \Delta} = 5\times 10^{-28} \rm{cm}^2 $, is the
resonant proton photon cross section for $\Delta$ production. Eqn. (\ref{phnd})
implies at a distance $r$ from the center, 
$n_{\gamma} = \frac{L_{iso}}{4 \pi r^2 \Gamma^2_h} T_{\gamma}$, where 
$T_{\gamma}= \epsilon_{\gamma}$, i.e., the energy of the photon. Using this 
relation, eqn.(\ref{rescond}) implies, 
\begin{eqnarray}
\frac{L_{iso} ~\sigma_{p \gamma \to \Delta}~ \Delta t }{ 4 \pi~ r^{2}_h 
\Gamma^{2}_h ~\epsilon_{\gamma}}= 1.
\label{rescond2}
\end{eqnarray}
In order to have resonance scattering, the energy condition, i.e.,
 eqn. (\ref{ph-pr-condn}) should be satisfied. Replacing $\epsilon_{\gamma}$ 
in terms of $\epsilon_p$ by using, (eqn.~(\ref{ph-pr-condn})~) and the 
expression for $\Gamma_h$ from eqn. (\ref{gammah}), one arrives at,
\begin{eqnarray}
r^{4}_h =\left[ \frac{ \sqrt{\left(R^{3} L_{iso} \rho_{*} \right)}~ 
\sigma_{p \gamma \to \Delta} ~\delta t ~\epsilon_p}
{ \sqrt{\pi} \times \left(.2 \mbox{~GeV}^2 \right)}\right]^{\frac{2}{5}}.
\end{eqnarray}
If we assume, the variability time $\delta t \sim 10^{-4}$ sec, 
$L_{iso}=10^{51} L_{52}$ erg/sec, and the energy of the proton $\epsilon_p=10^{5}$ GeV. Then
the radial distance from the center to have pion formation, $r_{\pi}$ turns 
out to be, $r_{\pi} \sim 10^{10.5}$ cm for $R=10^{12}$cm and $\rho_{*}= 10^{-7}
$ gm/cc.
\footnote{ We have used the fact $L_{iso}\delta t = 10^{51}L_{52} \rm{erg}= 
0.6 \times 10^{55} L_{52}$ GeV}. 
For more energetic protons, the radius of 
pion formation would shift further out-wards. For instance production of 
neutrinos of energy $\epsilon_{\nu} \ge 10^{17}$eV, would form further out.  
Once we have the estimate of the pion formation radius, we can estimate
the strength of the magnetic field and find out the field induced 
modification to the spectra. The estimation of the magnetic field is done 
in the next subsection.
%
%
\subsection{Magnetic Field Estimation}
%
%
In this note we have assumed the system to follow ideal MHD conditions.
That is to say that the magnetic energy density can be expressed  following
eqn. (\ref{cons-1})
\begin{eqnarray}
\left[ \frac{B^2_{o}}{2} \right] \sim 2.1 \cdot 10^{53} L^{3/4}_{52}.
\end{eqnarray}
Where we have used the energy density at the injection point i.e., 
$r=r_{o}=10^{7}$ cm with magnetic equipartition taken to be of order unity.
We note that  $r_o$ is also the width of the shell, that the GRB wind 
expands with. Conservation of magnetic field energy at the wind rest 
frame implies at any radial distance r, in the GRB environment,
\begin{eqnarray}
4 \pi r^2_o B^2_{o}= 4 \pi r^2 B^2.
\end{eqnarray}
leading to the relation, $B\times r = B_o \times r_o$ implying $B=B_o
\left( r_o/r \right)$. Using these relations, the magnetic field 
strength, in terms of GRB luminosity  turns out to be, 
$ B_o=\left[4.2 . 10^{53} L^{3/4}_{52}]^{1/2}\right] $. Hence the
magnetic field strength at distance r from the center turns out to be,
\begin{eqnarray}
B= \frac{r_o}{r}\left[4.2 . 10^{53} L^{3/4}_{52}\right]^{1/2} .
\label{magest}
\end{eqnarray}
For  conditions prevailing at the pion formation radius, corresponding to 
$10^{5}$ GeV protons, the strength of the field-- compared with the
critical Field strength $B_c (sim m^2_e)$-- i.e $\frac{B}{B_{cr}} \sim 0.01$.
The numbers mentioned here can get an order of magnitude variation depending
on the details. Since the purpose of this note is to get an order of magnitude
estimate of the effects, we reserve the details for future communications. 
\subsection{Estimates Of Energy Loss.}
The high energy $\pi^{+} \& \mu^{+}$ from the decay of $\Delta$ would undergo
energy loss by inverse Compton (IC) scattering, if the optical depth is large.
However even at low optical depth the end products remain coupled to photons
provided the photon drag time $t_D \sim \frac{m_p}{\sigma_T u_{\gamma}}$ is 
comparable to their decay time \cite{mlr}. One can verify that $t_D$ is indeed
comparable to the decay time, ensuring that the extremely relativistic
 pions and muons would undergo IC losses during their life time in the jet.
However during the same period, they are expected to be accelerated in the
reverse shocked plasma, to gain energy. Recalling the expression
of pressure at the shock front to be $P_r= 4 \left(  
\frac{\Gamma_j}{\Gamma_h}\right)^2 n_j m_p$, the equation of motion for the
charged particles at the shock front would be given by:
%
\begin{eqnarray}
\frac{dp}{dt}= 4 \left( \frac{\Gamma_j}{2\Gamma_h} \right)^2 n_j m_j 
\times~ Area.
\end{eqnarray}
%
If the jet makes an angle $\theta$, at a distance r from the center then 
the area would be $\pi (r\theta/2)^2$. One can further use the fact that
$\theta  \over \Gamma_j \sim 1$, one gets,
\begin{eqnarray}
\frac{dE_j}{E_j dt} = \frac{\pi n_j m_j r^2_o}{4 E_j} = \frac{\pi n_j r^2_o }{4\Gamma_j}.  
\end{eqnarray}
While coming to the relation above, we have made a conservative estimate, by 
replacing r by $r_o$ and neglecting the rest mass of the particles. During
the life time $\tau$, the fraction of energy gained, by the particles
 would be, 
\begin{eqnarray}
\Delta E_j = \frac{\pi n_j r^2_o E_j \tau}{4 \Gamma_j}.
\end{eqnarray}
The inverse Compton energy loss, in the ultra relativistic regime
 has already been estimated in \cite{raz-mes-wax}, and is given by
\begin{eqnarray}
\Delta_{IC}= \frac{E_j \tau \sigma_T (m_e/m_j)^2 n_{\gamma}}{\gamma_j 
(E_{\gamma}/m_j)}.
\end{eqnarray}
Finally the ratio,$\frac{\Delta_{IC}}{ \Delta_j} = \frac{4}{\pi}
\frac{\sigma_T m^2_e n_{\gamma}}{E_{\gamma}m_j n_j r^2_o}$. This ratio
probably would decide whether the IC loss is significant enough to cause
any significant suppression in neutrino production. For the kind of situation 
we are interested in, it seems this ratio would be less than unity hence, not 
much of suppression in the high energy neutrino spectrum. However this is a 
very tentative statement and one still needs to perform a quantitative 
estimation.
\section{Discussion and Outlook.}
%
High energy neutrinos moving in a magnetic field can decay to $\nu \to 
\nu l^{+} l^{-}$. Where $l$ can be either electron or muon or tau. The 
important point here is to note that, such processes are otherwise 
forbidden in vacuum. It has also been noted that, these processes are 
insensitive to neutrino mass and mixing in the leptonic sector. In a 
purely magnetic field the probablity of these processes depends on the 
parameter, $ {\cal{P}} =\frac{eBp_{\perp}}{m^3_{l}}$. Depending on whether
${\cal{P}}$ is greater than unity or less than unity, the probablity
for this reaction varies significantly. In a magnetized vacuum the 
probablity for such processes, $\alpha$ (for ${\cal{P}} >> 1 $), is given by
\begin{eqnarray}
\alpha_l=\frac{G^2_F\left(g^2_v+ g^2_A\right)m^6_l {{\cal{P}}}^2}{27 \pi^2 E_{\nu}}
\Bigg[ln {\cal{P}} -\frac{1}{2}ln3 -\gamma_E -\frac{29}{21} - 
\frac{1}{{{\cal{P}}^{\frac{2}{3}}}}
\frac{9}{56} \frac{3^{\frac{1}{3}}\pi^2}{\Gamma^4[\frac{2}{3}]} 
\frac{19g^2_V -63g^2_A}{g^2_V+ g^2_A} \Bigg].
\label{decay}
\end{eqnarray}
We would like to remind here that $\gamma_E$ is Euler's gamma function and 
$\Gamma[\frac{2}{3}]$ is the Gamma function. The last term in eqn. 
(\ref{decay}) contributes significantly when the neutrino flavor index and 
lepton flavor indices are different. we however would be interested in such 
a case.\\

\noindent
In a GRB, the location of the neutrino production of a particular flavor
strictly speaking would be different. However since the muon life time is    
pretty much small, therefore the location of $\nu$ production from $\pi^{+}$ 
decay and the same from $\mu^{+}$ decay though would be different however
the difference in length may not create significant changes in the medium
properties. Therefore, for all practical purposes, we would assume them to 
originate from the same location. If the number density of $\nu$ of certain 
energy $E_{\nu}$ at the production point is given by $n_{0\nu_{i}}(E)$ then
after travelling a distance dx the modified number density would be given by
$dn_{\nu_i}(E)= -n_{0\nu_i}(E) \alpha dx$. And the solution to 
the same eqn., gives $n_{\nu_i(E)}= n_{0\nu_i}(E) e^{-\alpha_{i} x}$; where 
$\alpha_i$ is the flavor dependent absorption coefficient. We can further
quantify the flavor dependent absorption as,
\begin{eqnarray}
\kappa_i=-ln\left( \frac{n_{\nu_i(E)}} {n_{0\nu_i}(E)}\right)= \alpha_i x.
\end{eqnarray}
Therefore one can define the asymmetry parameter $A_{p}$, to be:
\begin{eqnarray}
A_{p}= \frac{\kappa_{\nu_{\mu}} - \kappa_{\nu_e}}
{\kappa_{\nu_{\mu}}+\kappa_{\nu_e}} = \frac{\alpha_{\nu_{\mu}} - 
\alpha_{\nu_e}}{\alpha_{\nu_{\mu}}+\alpha_{\nu_e}}.
\end{eqnarray}
A measure of this quantity $A_{p}$ can in principle give us information
about the nature of the medium at the production site of the neutrinos 
as well as the neutrino properties.
In section ~[\ref{pir}]~ we had estimated the pion formation radius for
$\epsilon_p \ge 10^5GeV $ followed by the estimation of magnetic field 
at that location. 
\begin{figure}
  \begin{center}
     \input{rndecay.tex}
   \end{center}
\label{asymmetry}
\caption[]{Asymmetry parameter ($A_p$, see text) for $\nu_e$ and $\nu_{\mu}$ 
 for the High Energy neutrinos in the GRB enviornment. x - Axis corresponds
to neutrino energy in MeV and y Axis corresponds to $A_p$. Magnetic
field strength in units of electron mass is taken to be 0.001. Energy 
is taken to be same for both the flavors.}
\end{figure}
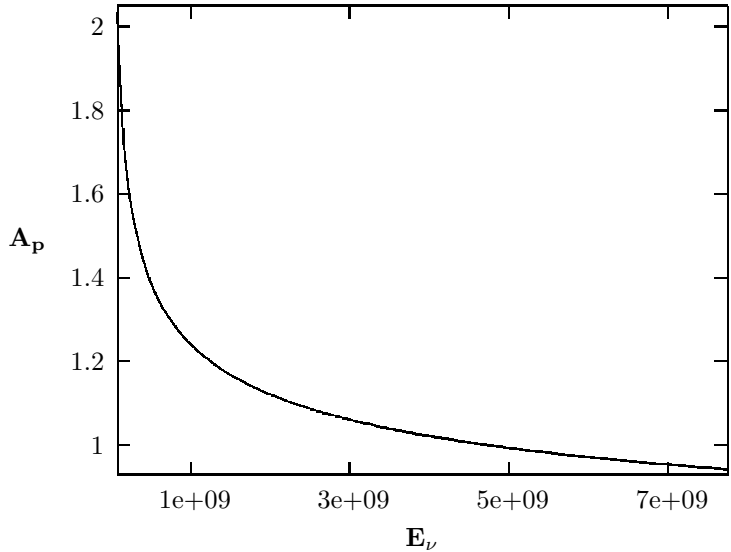 
%
It should be noted that any enhancement
of the magnetic fields due to charge particle streaming has been ignored here.
We assumed the magnetic field to be toroidal and the neutrino propagation
direction to be perpendicular to it. Furthermore we assumed the 
strength of the magnetic field to be held constant 
over the thickness of the shell (in its rest frame) i,e $r_o= 10^7$cm. This 
can be justified because any variation to magnetic field over a distance
larger than $\frac{1}{m_e}$ would not make any significant contribution
to the probablity amplitude we are interested in.\\

\noindent
Coming back to the question of asymmetry, as has already been discussed
the asymmetry parameter $A_p$ as has already been discussed, is far from 
zero. For very high energy neutrinos it seems to decrease
compared to the  low energy neutrinos, see fig.[\ref{asymmetry}]. However 
 estimation of actual degradation of energy spectrum of TeV or Super TeV
Neutrinos would require more careful estimation of the parameters discussed
here as well as  taking matter effects into consideration. However there
is an interesting consequence of this effect, that is, ther might 
be some enhancement in the number of neutrinos coming from the decay
of muons originating from the reaction $\nu_e \to \nu_e \mu^{+} \mu^{-}$.
That might have interesting consequences for the upcoming neutrino experiments.
%
%
\section{acknowledgement}
The author wishes to thank Profs. P.B. Pal and P. Bhattacharyya and K. Kar for 
discussions and exchange of views. He would also like to thank Dr. S. Razzaque
for bringing few articles to his notice.

\end{document}

%% file: rndecay.tex
\setlength{\unitlength}{0.240900pt}
\ifx\plotpoint\undefined\newsavebox{\plotpoint}\fi
\sbox{\plotpoint}{\rule[-0.200pt]{0.400pt}{0.400pt}}%
\begin{picture}(1200,900)(0,0)
\font\gnuplot=cmr10 at 10pt
\gnuplot
\sbox{\plotpoint}{\rule[-0.200pt]{0.400pt}{0.400pt}}%
\put(181.0,169.0){\rule[-0.200pt]{4.818pt}{0.400pt}}
\put(161,169){\makebox(0,0)[r]{ 1}}
\put(1119.0,169.0){\rule[-0.200pt]{4.818pt}{0.400pt}}
\put(181.0,301.0){\rule[-0.200pt]{4.818pt}{0.400pt}}
\put(161,301){\makebox(0,0)[r]{ 1.2}}
\put(1119.0,301.0){\rule[-0.200pt]{4.818pt}{0.400pt}}
\put(181.0,432.0){\rule[-0.200pt]{4.818pt}{0.400pt}}
\put(161,432){\makebox(0,0)[r]{ 1.4}}
\put(1119.0,432.0){\rule[-0.200pt]{4.818pt}{0.400pt}}
\put(181.0,564.0){\rule[-0.200pt]{4.818pt}{0.400pt}}
\put(161,564){\makebox(0,0)[r]{ 1.6}}
\put(1119.0,564.0){\rule[-0.200pt]{4.818pt}{0.400pt}}
\put(181.0,695.0){\rule[-0.200pt]{4.818pt}{0.400pt}}
\put(161,695){\makebox(0,0)[r]{ 1.8}}
\put(1119.0,695.0){\rule[-0.200pt]{4.818pt}{0.400pt}}
\put(181.0,827.0){\rule[-0.200pt]{4.818pt}{0.400pt}}
\put(161,827){\makebox(0,0)[r]{ 2}}
\put(1119.0,827.0){\rule[-0.200pt]{4.818pt}{0.400pt}}
\put(296.0,123.0){\rule[-0.200pt]{0.400pt}{4.818pt}}
\put(296,82){\makebox(0,0){ 1e+09}}
\put(296.0,840.0){\rule[-0.200pt]{0.400pt}{4.818pt}}
\put(545.0,123.0){\rule[-0.200pt]{0.400pt}{4.818pt}}
\put(545,82){\makebox(0,0){ 3e+09}}
\put(545.0,840.0){\rule[-0.200pt]{0.400pt}{4.818pt}}
\put(795.0,123.0){\rule[-0.200pt]{0.400pt}{4.818pt}}
\put(795,82){\makebox(0,0){ 5e+09}}
\put(795.0,840.0){\rule[-0.200pt]{0.400pt}{4.818pt}}
\put(1045.0,123.0){\rule[-0.200pt]{0.400pt}{4.818pt}}
\put(1045,82){\makebox(0,0){ 7e+09}}
\put(1045.0,840.0){\rule[-0.200pt]{0.400pt}{4.818pt}}
\put(181.0,123.0){\rule[-0.200pt]{230.782pt}{0.400pt}}
\put(1139.0,123.0){\rule[-0.200pt]{0.400pt}{177.543pt}}
\put(181.0,860.0){\rule[-0.200pt]{230.782pt}{0.400pt}}
\put(40,491){\makebox(0,0){${\bf{A_p}}$}}
\put(660,21){\makebox(0,0){$\bf{E_{\nu}}$}}
\put(181.0,123.0){\rule[-0.200pt]{0.400pt}{177.543pt}}
\put(181,849){\usebox{\plotpoint}}
\multiput(181.58,815.38)(0.491,-10.399){17}{\rule{0.118pt}{8.100pt}}
\multiput(180.17,832.19)(10.000,-183.188){2}{\rule{0.400pt}{4.050pt}}
\multiput(191.58,633.31)(0.491,-4.769){17}{\rule{0.118pt}{3.780pt}}
\multiput(190.17,641.15)(10.000,-84.154){2}{\rule{0.400pt}{1.890pt}}
\multiput(201.58,548.13)(0.492,-2.618){19}{\rule{0.118pt}{2.136pt}}
\multiput(200.17,552.57)(11.000,-51.566){2}{\rule{0.400pt}{1.068pt}}
\multiput(212.58,493.94)(0.491,-2.059){17}{\rule{0.118pt}{1.700pt}}
\multiput(211.17,497.47)(10.000,-36.472){2}{\rule{0.400pt}{0.850pt}}
\multiput(222.58,455.44)(0.491,-1.590){17}{\rule{0.118pt}{1.340pt}}
\multiput(221.17,458.22)(10.000,-28.219){2}{\rule{0.400pt}{0.670pt}}
\multiput(232.58,425.60)(0.491,-1.225){17}{\rule{0.118pt}{1.060pt}}
\multiput(231.17,427.80)(10.000,-21.800){2}{\rule{0.400pt}{0.530pt}}
\multiput(242.58,402.26)(0.491,-1.017){17}{\rule{0.118pt}{0.900pt}}
\multiput(241.17,404.13)(10.000,-18.132){2}{\rule{0.400pt}{0.450pt}}
\multiput(252.58,383.17)(0.492,-0.732){19}{\rule{0.118pt}{0.682pt}}
\multiput(251.17,384.58)(11.000,-14.585){2}{\rule{0.400pt}{0.341pt}}
\multiput(263.58,367.09)(0.491,-0.756){17}{\rule{0.118pt}{0.700pt}}
\multiput(262.17,368.55)(10.000,-13.547){2}{\rule{0.400pt}{0.350pt}}
\multiput(273.58,352.43)(0.491,-0.652){17}{\rule{0.118pt}{0.620pt}}
\multiput(272.17,353.71)(10.000,-11.713){2}{\rule{0.400pt}{0.310pt}}
\multiput(283.58,339.76)(0.491,-0.547){17}{\rule{0.118pt}{0.540pt}}
\multiput(282.17,340.88)(10.000,-9.879){2}{\rule{0.400pt}{0.270pt}}
\multiput(293.00,329.92)(0.495,-0.491){17}{\rule{0.500pt}{0.118pt}}
\multiput(293.00,330.17)(8.962,-10.000){2}{\rule{0.250pt}{0.400pt}}
\multiput(303.00,319.93)(0.553,-0.489){15}{\rule{0.544pt}{0.118pt}}
\multiput(303.00,320.17)(8.870,-9.000){2}{\rule{0.272pt}{0.400pt}}
\multiput(313.00,310.93)(0.692,-0.488){13}{\rule{0.650pt}{0.117pt}}
\multiput(313.00,311.17)(9.651,-8.000){2}{\rule{0.325pt}{0.400pt}}
\multiput(324.00,302.93)(0.626,-0.488){13}{\rule{0.600pt}{0.117pt}}
\multiput(324.00,303.17)(8.755,-8.000){2}{\rule{0.300pt}{0.400pt}}
\multiput(334.00,294.93)(0.721,-0.485){11}{\rule{0.671pt}{0.117pt}}
\multiput(334.00,295.17)(8.606,-7.000){2}{\rule{0.336pt}{0.400pt}}
\multiput(344.00,287.93)(0.721,-0.485){11}{\rule{0.671pt}{0.117pt}}
\multiput(344.00,288.17)(8.606,-7.000){2}{\rule{0.336pt}{0.400pt}}
\multiput(354.00,280.93)(0.852,-0.482){9}{\rule{0.767pt}{0.116pt}}
\multiput(354.00,281.17)(8.409,-6.000){2}{\rule{0.383pt}{0.400pt}}
\multiput(364.00,274.93)(1.155,-0.477){7}{\rule{0.980pt}{0.115pt}}
\multiput(364.00,275.17)(8.966,-5.000){2}{\rule{0.490pt}{0.400pt}}
\multiput(375.00,269.93)(0.852,-0.482){9}{\rule{0.767pt}{0.116pt}}
\multiput(375.00,270.17)(8.409,-6.000){2}{\rule{0.383pt}{0.400pt}}
\multiput(385.00,263.93)(1.044,-0.477){7}{\rule{0.900pt}{0.115pt}}
\multiput(385.00,264.17)(8.132,-5.000){2}{\rule{0.450pt}{0.400pt}}
\multiput(395.00,258.94)(1.358,-0.468){5}{\rule{1.100pt}{0.113pt}}
\multiput(395.00,259.17)(7.717,-4.000){2}{\rule{0.550pt}{0.400pt}}
\multiput(405.00,254.93)(1.044,-0.477){7}{\rule{0.900pt}{0.115pt}}
\multiput(405.00,255.17)(8.132,-5.000){2}{\rule{0.450pt}{0.400pt}}
\multiput(415.00,249.94)(1.505,-0.468){5}{\rule{1.200pt}{0.113pt}}
\multiput(415.00,250.17)(8.509,-4.000){2}{\rule{0.600pt}{0.400pt}}
\multiput(426.00,245.94)(1.358,-0.468){5}{\rule{1.100pt}{0.113pt}}
\multiput(426.00,246.17)(7.717,-4.000){2}{\rule{0.550pt}{0.400pt}}
\multiput(436.00,241.94)(1.358,-0.468){5}{\rule{1.100pt}{0.113pt}}
\multiput(436.00,242.17)(7.717,-4.000){2}{\rule{0.550pt}{0.400pt}}
\multiput(446.00,237.94)(1.358,-0.468){5}{\rule{1.100pt}{0.113pt}}
\multiput(446.00,238.17)(7.717,-4.000){2}{\rule{0.550pt}{0.400pt}}
\multiput(456.00,233.95)(2.025,-0.447){3}{\rule{1.433pt}{0.108pt}}
\multiput(456.00,234.17)(7.025,-3.000){2}{\rule{0.717pt}{0.400pt}}
\multiput(466.00,230.95)(2.248,-0.447){3}{\rule{1.567pt}{0.108pt}}
\multiput(466.00,231.17)(7.748,-3.000){2}{\rule{0.783pt}{0.400pt}}
\multiput(477.00,227.94)(1.358,-0.468){5}{\rule{1.100pt}{0.113pt}}
\multiput(477.00,228.17)(7.717,-4.000){2}{\rule{0.550pt}{0.400pt}}
\multiput(487.00,223.95)(2.025,-0.447){3}{\rule{1.433pt}{0.108pt}}
\multiput(487.00,224.17)(7.025,-3.000){2}{\rule{0.717pt}{0.400pt}}
\multiput(497.00,220.95)(2.025,-0.447){3}{\rule{1.433pt}{0.108pt}}
\multiput(497.00,221.17)(7.025,-3.000){2}{\rule{0.717pt}{0.400pt}}
\put(507,217.17){\rule{2.100pt}{0.400pt}}
\multiput(507.00,218.17)(5.641,-2.000){2}{\rule{1.050pt}{0.400pt}}
\multiput(517.00,215.95)(2.248,-0.447){3}{\rule{1.567pt}{0.108pt}}
\multiput(517.00,216.17)(7.748,-3.000){2}{\rule{0.783pt}{0.400pt}}
\multiput(528.00,212.95)(2.025,-0.447){3}{\rule{1.433pt}{0.108pt}}
\multiput(528.00,213.17)(7.025,-3.000){2}{\rule{0.717pt}{0.400pt}}
\put(538,209.17){\rule{2.100pt}{0.400pt}}
\multiput(538.00,210.17)(5.641,-2.000){2}{\rule{1.050pt}{0.400pt}}
\multiput(548.00,207.95)(2.025,-0.447){3}{\rule{1.433pt}{0.108pt}}
\multiput(548.00,208.17)(7.025,-3.000){2}{\rule{0.717pt}{0.400pt}}
\put(558,204.17){\rule{2.100pt}{0.400pt}}
\multiput(558.00,205.17)(5.641,-2.000){2}{\rule{1.050pt}{0.400pt}}
\put(568,202.17){\rule{2.100pt}{0.400pt}}
\multiput(568.00,203.17)(5.641,-2.000){2}{\rule{1.050pt}{0.400pt}}
\multiput(578.00,200.95)(2.248,-0.447){3}{\rule{1.567pt}{0.108pt}}
\multiput(578.00,201.17)(7.748,-3.000){2}{\rule{0.783pt}{0.400pt}}
\put(589,197.17){\rule{2.100pt}{0.400pt}}
\multiput(589.00,198.17)(5.641,-2.000){2}{\rule{1.050pt}{0.400pt}}
\put(599,195.17){\rule{2.100pt}{0.400pt}}
\multiput(599.00,196.17)(5.641,-2.000){2}{\rule{1.050pt}{0.400pt}}
\put(609,193.17){\rule{2.100pt}{0.400pt}}
\multiput(609.00,194.17)(5.641,-2.000){2}{\rule{1.050pt}{0.400pt}}
\put(619,191.17){\rule{2.100pt}{0.400pt}}
\multiput(619.00,192.17)(5.641,-2.000){2}{\rule{1.050pt}{0.400pt}}
\put(629,189.17){\rule{2.300pt}{0.400pt}}
\multiput(629.00,190.17)(6.226,-2.000){2}{\rule{1.150pt}{0.400pt}}
\put(640,187.17){\rule{2.100pt}{0.400pt}}
\multiput(640.00,188.17)(5.641,-2.000){2}{\rule{1.050pt}{0.400pt}}
\put(650,185.17){\rule{2.100pt}{0.400pt}}
\multiput(650.00,186.17)(5.641,-2.000){2}{\rule{1.050pt}{0.400pt}}
\put(660,183.67){\rule{2.409pt}{0.400pt}}
\multiput(660.00,184.17)(5.000,-1.000){2}{\rule{1.204pt}{0.400pt}}
\put(670,182.17){\rule{2.100pt}{0.400pt}}
\multiput(670.00,183.17)(5.641,-2.000){2}{\rule{1.050pt}{0.400pt}}
\put(680,180.17){\rule{2.300pt}{0.400pt}}
\multiput(680.00,181.17)(6.226,-2.000){2}{\rule{1.150pt}{0.400pt}}
\put(691,178.67){\rule{2.409pt}{0.400pt}}
\multiput(691.00,179.17)(5.000,-1.000){2}{\rule{1.204pt}{0.400pt}}
\put(701,177.17){\rule{2.100pt}{0.400pt}}
\multiput(701.00,178.17)(5.641,-2.000){2}{\rule{1.050pt}{0.400pt}}
\put(711,175.17){\rule{2.100pt}{0.400pt}}
\multiput(711.00,176.17)(5.641,-2.000){2}{\rule{1.050pt}{0.400pt}}
\put(721,173.67){\rule{2.409pt}{0.400pt}}
\multiput(721.00,174.17)(5.000,-1.000){2}{\rule{1.204pt}{0.400pt}}
\put(731,172.17){\rule{2.300pt}{0.400pt}}
\multiput(731.00,173.17)(6.226,-2.000){2}{\rule{1.150pt}{0.400pt}}
\put(742,170.67){\rule{2.409pt}{0.400pt}}
\multiput(742.00,171.17)(5.000,-1.000){2}{\rule{1.204pt}{0.400pt}}
\put(752,169.17){\rule{2.100pt}{0.400pt}}
\multiput(752.00,170.17)(5.641,-2.000){2}{\rule{1.050pt}{0.400pt}}
\put(762,167.67){\rule{2.409pt}{0.400pt}}
\multiput(762.00,168.17)(5.000,-1.000){2}{\rule{1.204pt}{0.400pt}}
\put(772,166.67){\rule{2.409pt}{0.400pt}}
\multiput(772.00,167.17)(5.000,-1.000){2}{\rule{1.204pt}{0.400pt}}
\put(782,165.17){\rule{2.100pt}{0.400pt}}
\multiput(782.00,166.17)(5.641,-2.000){2}{\rule{1.050pt}{0.400pt}}
\put(792,163.67){\rule{2.650pt}{0.400pt}}
\multiput(792.00,164.17)(5.500,-1.000){2}{\rule{1.325pt}{0.400pt}}
\put(803,162.67){\rule{2.409pt}{0.400pt}}
\multiput(803.00,163.17)(5.000,-1.000){2}{\rule{1.204pt}{0.400pt}}
\put(813,161.17){\rule{2.100pt}{0.400pt}}
\multiput(813.00,162.17)(5.641,-2.000){2}{\rule{1.050pt}{0.400pt}}
\put(823,159.67){\rule{2.409pt}{0.400pt}}
\multiput(823.00,160.17)(5.000,-1.000){2}{\rule{1.204pt}{0.400pt}}
\put(833,158.67){\rule{2.409pt}{0.400pt}}
\multiput(833.00,159.17)(5.000,-1.000){2}{\rule{1.204pt}{0.400pt}}
\put(843,157.67){\rule{2.650pt}{0.400pt}}
\multiput(843.00,158.17)(5.500,-1.000){2}{\rule{1.325pt}{0.400pt}}
\put(854,156.67){\rule{2.409pt}{0.400pt}}
\multiput(854.00,157.17)(5.000,-1.000){2}{\rule{1.204pt}{0.400pt}}
\put(864,155.17){\rule{2.100pt}{0.400pt}}
\multiput(864.00,156.17)(5.641,-2.000){2}{\rule{1.050pt}{0.400pt}}
\put(874,153.67){\rule{2.409pt}{0.400pt}}
\multiput(874.00,154.17)(5.000,-1.000){2}{\rule{1.204pt}{0.400pt}}
\put(884,152.67){\rule{2.409pt}{0.400pt}}
\multiput(884.00,153.17)(5.000,-1.000){2}{\rule{1.204pt}{0.400pt}}
\put(894,151.67){\rule{2.650pt}{0.400pt}}
\multiput(894.00,152.17)(5.500,-1.000){2}{\rule{1.325pt}{0.400pt}}
\put(905,150.67){\rule{2.409pt}{0.400pt}}
\multiput(905.00,151.17)(5.000,-1.000){2}{\rule{1.204pt}{0.400pt}}
\put(915,149.67){\rule{2.409pt}{0.400pt}}
\multiput(915.00,150.17)(5.000,-1.000){2}{\rule{1.204pt}{0.400pt}}
\put(925,148.67){\rule{2.409pt}{0.400pt}}
\multiput(925.00,149.17)(5.000,-1.000){2}{\rule{1.204pt}{0.400pt}}
\put(935,147.67){\rule{2.409pt}{0.400pt}}
\multiput(935.00,148.17)(5.000,-1.000){2}{\rule{1.204pt}{0.400pt}}
\put(945,146.67){\rule{2.650pt}{0.400pt}}
\multiput(945.00,147.17)(5.500,-1.000){2}{\rule{1.325pt}{0.400pt}}
\put(956,145.67){\rule{2.409pt}{0.400pt}}
\multiput(956.00,146.17)(5.000,-1.000){2}{\rule{1.204pt}{0.400pt}}
\put(966,144.67){\rule{2.409pt}{0.400pt}}
\multiput(966.00,145.17)(5.000,-1.000){2}{\rule{1.204pt}{0.400pt}}
\put(976,143.67){\rule{2.409pt}{0.400pt}}
\multiput(976.00,144.17)(5.000,-1.000){2}{\rule{1.204pt}{0.400pt}}
\put(986,142.67){\rule{2.409pt}{0.400pt}}
\multiput(986.00,143.17)(5.000,-1.000){2}{\rule{1.204pt}{0.400pt}}
\put(996,141.67){\rule{2.650pt}{0.400pt}}
\multiput(996.00,142.17)(5.500,-1.000){2}{\rule{1.325pt}{0.400pt}}
\put(1007,140.67){\rule{2.409pt}{0.400pt}}
\multiput(1007.00,141.17)(5.000,-1.000){2}{\rule{1.204pt}{0.400pt}}
\put(1017,139.67){\rule{2.409pt}{0.400pt}}
\multiput(1017.00,140.17)(5.000,-1.000){2}{\rule{1.204pt}{0.400pt}}
\put(1027,138.67){\rule{2.409pt}{0.400pt}}
\multiput(1027.00,139.17)(5.000,-1.000){2}{\rule{1.204pt}{0.400pt}}
\put(1047,137.67){\rule{2.409pt}{0.400pt}}
\multiput(1047.00,138.17)(5.000,-1.000){2}{\rule{1.204pt}{0.400pt}}
\put(1057,136.67){\rule{2.650pt}{0.400pt}}
\multiput(1057.00,137.17)(5.500,-1.000){2}{\rule{1.325pt}{0.400pt}}
\put(1068,135.67){\rule{2.409pt}{0.400pt}}
\multiput(1068.00,136.17)(5.000,-1.000){2}{\rule{1.204pt}{0.400pt}}
\put(1078,134.67){\rule{2.409pt}{0.400pt}}
\multiput(1078.00,135.17)(5.000,-1.000){2}{\rule{1.204pt}{0.400pt}}
\put(1088,133.67){\rule{2.409pt}{0.400pt}}
\multiput(1088.00,134.17)(5.000,-1.000){2}{\rule{1.204pt}{0.400pt}}
\put(1037.0,139.0){\rule[-0.200pt]{2.409pt}{0.400pt}}
\put(1108,132.67){\rule{2.650pt}{0.400pt}}
\multiput(1108.00,133.17)(5.500,-1.000){2}{\rule{1.325pt}{0.400pt}}
\put(1119,131.67){\rule{2.409pt}{0.400pt}}
\multiput(1119.00,132.17)(5.000,-1.000){2}{\rule{1.204pt}{0.400pt}}
\put(1129,130.67){\rule{2.409pt}{0.400pt}}
\multiput(1129.00,131.17)(5.000,-1.000){2}{\rule{1.204pt}{0.400pt}}
\put(1098.0,134.0){\rule[-0.200pt]{2.409pt}{0.400pt}}
\put(1139,131){\usebox{\plotpoint}}
\end{picture}

%% file: rmag-new-grb.bbl
\begin{thebibliography}{[9]}
%



 





\bibitem{compact} Gulibert, P.W., Fabian, A,C., Rees, M.J., MNRS, {\bf{205}}
593(1983). Belinnikov, S., astroph/9911138.

\bibitem{mes}M.J. Rees and  P. Meszaros , Astrophys. J.430, L93 (1994). 




\bibitem{ganguly} K. Bhattacharya, A. K. Ganguly and S. Konar, Phys. Rev. 
{\bf{D65}} 013007,2002. K. Bhattacharya and Avijit K. Ganguly Phys. Rev. 
 {\bf{D68}} 053011 (2003). K. Bhattacharya and Avijit. K. Ganguly
:hep-ph/0209237; and see the references there in. S.P. Mikheyev and 
A. Au.Smirnov, Sov.J. Nucl. Phys. For a detailed exposition
to this topic one can see, K. Bhattacharyya and P.~B.~Pal. Proc.
Indian. Natn Sci Acad. {\bf{70 A}} No 1. January 2004, pp. 145-161.

{\bf{42}}, 913 (1985). L. Wolfenstein, Phys. Rev {\bf{D17}}, 2369 (1978).
 

\bibitem{bet} H. A. Bethe,  Rev. Of. Mod. Phys. {\bf{62}}, 1990, 801. 

\bibitem{bah-mes} J. N. Bahcall and P. Me'sza'ros, Phys. Rev. Lett. 85,
1362 (2000).


\bibitem{wax-bah} E. Waxman and J. Bahcall, Phys. Rev. Lett. 78, 2292
(1997).  

\bibitem{wax-mes} P. Meszaros and  E. Waxman and , Phys. Rev. Lett. 87,
171102-1 (2001). 

\bibitem{mackee} J. Tan, C. D. Matzner and C.F. McKee, Astrophys. {\bf{J}}
946 (2001).

\bibitem{waxmanl} Eli Waxman, astroph/0303517.

\bibitem{rees} M.~J. Rees and P. Meszaros, Astrophys. J. Lett, 430, L93.

\bibitem{waxr} see for instance reference  \cite{waxmanl}

\bibitem{raz-mes-wax} Soebur Razzaque, Peter Me'sza'ros and Eli Waxman,
astroph/0303505

\bibitem{mlr} P. Meszaros, P. Laguna and M. J. Rees, astro-ph/9301007.



\bibitem{kuz} A.~V~. Kuznetsov, N.~V.~Mikheev, Phys. Lett. {\bf B394 },123 
(197).



\end{thebibliography}
